\documentclass[prl,english,twocolumn,superscriptaddress,showpacs,floatfix,amsfonts]{revtex4}
\usepackage{amssymb}
\usepackage{amsmath}
\usepackage{amsthm}
\usepackage{color}
\usepackage{yhmath}
\usepackage{cancel}
\usepackage{graphicx}%
\usepackage{epstopdf}
\makeatletter
\usepackage{babel}
\makeatother
\begin{document}
\title{Kondo Resonance in the Presence of Spin-Polarized Currents}
\author{Yunong Qi}
\affiliation{Texas Center for Superconductivity, University of Houston, Houston, Texas 77204}
\author{Jian-Xin Zhu}
\email{jxzhu@lanl.gov}
\homepage{http://theory.lanl.gov}
\affiliation{Theoretical Division, Los Alamos National Laboratory, Los Alamos, New Mexico 87545}
\author{Shufeng Zhang}
\affiliation{Department of Physics
\& Astronomy, University of Missouri, Columbia, Missouri 65211}
\author{C. S. Ting}
\email{csting@mail.uh.edu}
\affiliation{Texas Center for Superconductivity, University of Houston, Houston, Texas 77204}

\date{\today}
\begin{abstract}
We propose an improved method of the equation of motion approach to study
the Kondo problem in spin-dependent non-equilibrium conditions. We find that
the previously introduced additional renormalization for non-equilibrium
Kondo effects is not required when we use a proper decoupling scheme. Our 
improved formulation is then applied to address the spin-split Kondo peaks
when a spin current injects into a Kondo system.

\end{abstract}
\maketitle

The Kondo effect~\cite{Hewson} has been one of the central and challenging 
problems in condensed matter physics for many years. In equilibrium, the Kondo
peak formed at the Fermi level is degenerate with respect to
spin degrees of freedom when a magnetic impurity is embedded in a non-magnetic metal. 
By applying a magnetic field, the spin degeneracy of the Kondo peak
can be lifted due to Zeeman splitting of the impurity level. For a mesoscopic
system such as a quantum dot, one can also use ferromagnetic leads to lift
the spin degeneracy; this is because the ferromagnetic leads provide the spin
dependence of the interaction (or hybridization) between the conduction 
electrons of the leads and the localized state in the quantum dot. 
To explain the above spin-dependent non-equilibrium phenomena, one frequently 
relies on the method of the equation motion (EOM) where an additional
renormalization in the self-energy is introduced without rigorous 
justification.

In this Letter, we address two issues. First, we show that the 
non-equilibrium Kondo problem can be solved without the
artificial additional renormalization as long as one correctly keeps
the previously neglected terms in the EOM. We establish a new formulation
for the case of finite and infinite $U$ in the Anderson model. Second, we
use the improved EOM method to predict spin-split Kondo peaks in the
presence of a non-equilibrium spin accumulation. 
The spin accumulation has played a crucial role in
the emerging field of spin electronics ~\cite{Wolf, Fiederling, Kane}. 
Our prediction directly connects the separation of the Kondo peaks with
the spin accumulation, and thus it provides a way to determine the spin
accumulation in Kondo systems.
We emphasize an important distinction of the present study 
from previous one: the non-equilibrium condition generated
by a voltage across the quantum dot and the ferromagnetic lead has a common 
chemical potential for spin up and down electrons, while 
the spin accumulation generates spin-split chemical potentials.
 
There are a number of methods to study the Kondo effect in the 
spin-dependent non-equilibrium condition. The EOM approach of 
the Anderson model has been used for treating both equilibrium and 
non-equilibrium Kondo physics at low temperatures. The EOM approach 
includes re-summation of low-order hopping processes and needs a decoupling 
scheme in order to obtain a closed analytical form. We follow the procedure
introduced by Appelbaum, Penn, and Lacroix (APL)~\cite{Appelbaum,Lacroix}, 
which is known to capture the right qualitative feature of physics at low 
temperatures. We note that other approaches, e.g., numerical renormalization group method, 
can also describe the low temperature Kondo effect \cite{Martinek2, Choi}. 

The Hamiltonian of the impurity Anderson model is 
\begin{align}
H =
\displaystyle\sum_{k,{\sigma}}{\epsilon}_{k{\sigma}}c^{\dagger}_{k{\sigma}}c_{k{\sigma}}
+ \displaystyle\sum_{\sigma}{\epsilon}_{d{\sigma}}d^{\dagger}_{\sigma}d_{\sigma} +
Ud^{\dagger}_{\sigma}d_{\sigma}d^{\dagger}_{\bar{\sigma}}d_{\bar{\sigma}} 
\nonumber
\\
+
\displaystyle\sum_{k,{\sigma}}V_{k{\sigma}}\left[c^{\dagger}_{k{\sigma}}d_{\sigma} +
d^{\dagger}_{\sigma}c_{k{\sigma}}\right] \;.
\end{align}
Here $c^{\dagger}_{k \sigma}$ and $d^{\dagger}_{\sigma}$ are respectively the 
creation operators for conduction and $d$ electrons at the impurity site. 
The quantities ${\epsilon}_{k{\sigma}}$, ${\epsilon}_{d\sigma}$ are the 
conduction electron energy dispersion and the impurity energy, respectively. 
We assume that conduction electrons density of states is 
constant but spin dependent, i.e., 
$\rho(\epsilon_{\sigma})=1/2D_{\sigma}$ when 
$-D_{\sigma} {\le} {\epsilon}_{k{\sigma}}{\le} D_{\sigma}$, 
$U$ is the intra-atomic Coulomb interaction at the impurity site, and 
$V_{k\sigma}$ represents the $s$-$d$ hybridization.

By using the standard procedure for the EOM, we 
obtain the impurity Green function, 
$G_{d\sigma}\stackrel{\text{\tiny def}}{=}{\ll}d_{\sigma}{\mid}d^{\dagger}_
{\sigma}{\gg}$, 
\begin{align}
G_{d\sigma} = \dfrac{1 - \bar{n}_{d\bar{\sigma}}\left(\omega\right) }
{\omega - {\epsilon}_{d\sigma} - {\Sigma}_{0\sigma} + 
\dfrac{ U {\Sigma}_{1\sigma}}{\omega - {\epsilon}_{d\sigma} - 
U - {\Sigma}_{0\sigma} - {\Sigma}_{3\sigma}}} 
\nonumber
\\
+ \dfrac{\bar{n}_{d\bar{\sigma}}}{\omega - {\epsilon}_{d\sigma} - 
{\Sigma}_{0\sigma}  -U - \dfrac{U\left[{\Sigma}_{3\sigma} - 
{\Sigma}_{1\sigma}\right] }{\omega - {\epsilon}_{d\sigma} - 
U - {\Sigma}_{0\sigma} - {\Sigma}_{3\sigma}}}\;.  
\label{EQ:dGreen}
\end{align}
To arrive at  Eq.~(\ref{EQ:dGreen}), we have made the decoupling approximation 
procedure shown below. The ``dynamic''  
average occupation number (i.e., frequency-dependence) of the impurity level
is defined as
\begin{equation}
\bar{n}_{d\bar{\sigma}} \left({\omega}\right)  \stackrel{\text{\tiny def}}{=} 
{\langle}n_{d\bar{\sigma}}{\rangle}  - 
\displaystyle\sum_{q}\dfrac{V_{q\bar{\sigma}}{
\langle}c^{\dagger}_{q\bar{\sigma}}d_{\bar{\sigma}}{\rangle}}
{D_{1\sigma}\left(\omega,q\right)} + \displaystyle\sum_{q}\dfrac{V_{q
\bar{\sigma}}^{*}{\langle}d^{\dagger}_{\bar{\sigma}}c_{q\bar{\sigma}}
{\rangle}}{D_{2\sigma}\left(\omega,q\right)} \;,
\end{equation}
and the three self-energies are
\begin{align}
{\Sigma}_{0\sigma}  \stackrel{\text{\tiny def}}{=} \displaystyle\sum_{k}
\dfrac{\vert V_{k\sigma}\vert^{2}}{{\omega} - {\epsilon}_{k\sigma}}\;,
\end{align}
\begin{align}
{\Sigma}_{1\sigma}  \stackrel{\text{\tiny def}}{=} \displaystyle\sum_{k} \displaystyle\sum_{q}
\left[\dfrac{V_{k\bar{\sigma}}^{*}V_{q\bar{\sigma}}{\langle}c^{\dagger}_{q\bar{\sigma}}c_{k\bar{\sigma}}{\rangle}}{D_{2\sigma}\left(\omega, k\right)}  +  \dfrac{V_{k\bar{\sigma}}V_{q\bar{\sigma}}^{*}{\langle}c^{\dagger}_{k\bar{\sigma}}c_{q\bar{\sigma}}{\rangle}}{D_{1\sigma}\left(\omega,k\right)} \right] 
\nonumber
\\
+   \displaystyle\sum_{k}\left[\dfrac{V_{k\bar{\sigma}}^{*}{\langle}c_{k\bar{\sigma}}d^{\dagger}_{\bar{\sigma}}{\rangle}}{D_{2\sigma}\left({\omega},k \right)} + \dfrac{V_{k\bar{\sigma}}{\langle}
c^{\dagger}_{k\bar{\sigma}}d_{\bar{\sigma}}{\rangle}}{D_{1\sigma}\left({\omega},k\right)}\right] \Sigma_{0\sigma} \;,
\label{EQ:Sigma1}
\end{align}
and
\begin{align}
{\Sigma}_{3\sigma}  \stackrel{\text{\tiny def}}{=}\displaystyle\sum_{k}\dfrac{\vert V_{k\bar{\sigma}}\vert^{2}}{D_{1\sigma}\left(\omega,k \right)} + \displaystyle\sum_{k}\dfrac{\vert V_{k\bar{\sigma}}\vert^{2}}{D_{2\sigma}\left(\omega,k \right)} \;,
\end{align}      
where 
 \begin{align}
 D_{1\sigma}\left(\omega,k\right)  \stackrel{\text{\tiny def}}{=} {\omega} + {\epsilon}_{k\bar{\sigma}} - {\epsilon}_{d\sigma} - {\epsilon}_{d\bar{\sigma}} - U\;,
 \end{align}
and
 \begin{align}
 D_{2\sigma}\left(\omega,k\right)  \stackrel{\text{\tiny def}}{=} {\omega} - {\epsilon}_{k\bar{\sigma}} - {\epsilon}_{d\sigma} + {\epsilon}_{d\bar{\sigma}}. 
 \end{align}
We shall point out that the expectation value of ${\langle}
c^{\dagger}_{q\bar{\sigma}}c_{k\bar{\sigma}}{\rangle}$, 
${\langle}d^{\dagger}_{\bar{\sigma}}c_{k\bar{\sigma}}{\rangle}$ which have 
been discarded in previous EOM studies are important 
at low temperatures since they diverge logarithmically at the Fermi level 
as the temperature approaches to zero. Their values should be self-consistently
evaluated through the following identities:   
\begin{equation}
 {\langle}c^{\dagger}_{q\bar{\sigma}}c_{k\bar{\sigma}}{\rangle} =
- \dfrac{1}{\pi}{\displaystyle\int}f_{FD}(\omega)\mbox{Im}{\ll}c_{k\bar{\sigma}}{\mid}c^{\dagger}_{q\bar{\sigma}}{\gg}d{\omega}\;,
\end{equation}      
where $f_{FD}(\omega)= 1/[\it{e}^{{\beta}{\omega}} + 1]$ is the Fermi-Dirac distribution function, ${\beta} = 1/k_{B}T$, and
the Green function ${\ll}c_{k\bar{\sigma}}{\mid}c^{\dagger}_{q\bar{\sigma}}
{\gg}$ is:
\begin{equation}
{\ll}c_{k\bar{\sigma}}{\mid}c^{\dagger}_{q\bar{\sigma}}{\gg} =
\dfrac{{\delta}_{q,k}}{\omega - {\epsilon}_{k\bar{\sigma}}} +
\dfrac{V_{k\bar{\sigma}}V_{q\bar{\sigma}}^{*}{\ll}d_{\bar{\sigma}}{\mid}d^{\dagger}_{\bar{\sigma}}{\gg}}{\left(
\omega - {\epsilon}_{k\bar{\sigma}}\right)\left(\omega - {\epsilon}_{q\bar{\sigma}}\right)}\;,
\end{equation}
and similarly
\begin{equation}
{\langle}d^{\dagger}_{\bar{\sigma}}c_{k\bar{\sigma}}{\rangle} =
-\dfrac{1}{\pi}{\displaystyle\int}f(\omega)\mbox{Im}{\ll}c_{k\bar{\sigma}}{\mid}d^{\dagger}_{\bar{\sigma}}{\gg}d{\omega} \;,
\end{equation}
with 
\begin{equation}
{\ll}c_{k\bar{\sigma}}{\mid}d^{\dagger}_{\bar{\sigma}}{\gg} =  
\dfrac{V_{k\bar{\sigma}}{\ll}d_{\bar{\sigma}}{\mid}d^{\dagger}_{\bar{\sigma}}{\gg}}{\left(
\omega - {\epsilon}_{k\bar{\sigma}}\right)}\;.
\end{equation}

Equation (\ref{EQ:dGreen}) is an extension to a similar 
result obtained by Meir {\em et al.}~\cite{Meir91}. Equations (2) through (10)
constitute a set of the closed self-consistent equations which can be 
numerically solved. Before we carry out numerical calculations, we point out
the new ingredients in our formula: 1) the effective occupancy is 
frequency dependent, and 2) 
the higher order self-energy contains 
the intermediate off-diagonal states in momentum space 
(e.g., $\langle c_{k\sigma}^{\dagger}c_{q\sigma} \rangle$) and 
charge fluctuations (e.g., $\langle d_{\sigma}^{\dagger}c_{q\sigma} \rangle$).
Solving the coupled Eq.~(\ref{EQ:dGreen})-(10) not 
only yields the Kondo resonance at low temperatures but also allows us to 
explicitly include the logarithmic divergence in general. 
\begin{figure}[t]
\begin{center}
\includegraphics[width=8cm]{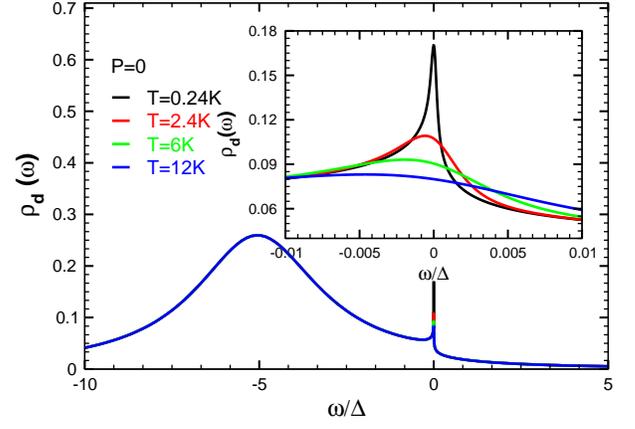}
\caption{(Color) Spectral density ${\rho}_{d\sigma}$ calculated via the EOM 
method for an infinity $U$ Anderson model for various temperatures 
in the absence of spin polarization. The inset displays the zoom-in 
view of the Kondo resonance near the Fermi energy.}
\label{fig:fig1.eps}
\end{center}
\end{figure}
\begin{figure}[t]
\begin{center}
\includegraphics[width=8cm]{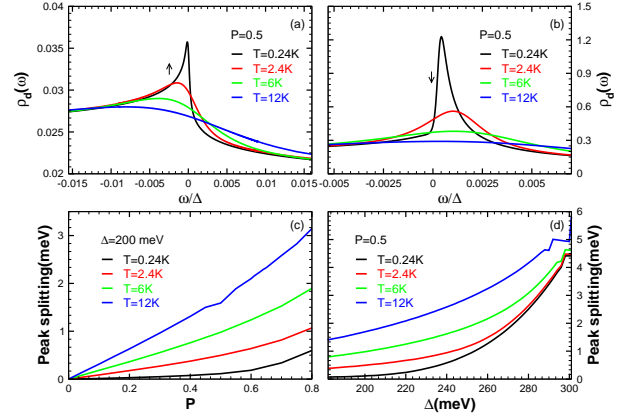}
\caption{ (Color) Spectral density ${\rho}_{d\sigma}$ calculated via the 
EOM method for an infinity $U$ Anderson model for various temperatures 
for a fixed degree, $P=0.5$, of spin polarization (a) and (b),
and the Kondo resonance splitting as a function of the degree of spin 
polarization (c) and the hybridization strength at a fixed $P=0.5$ (d).}
\label{fig:fig2.eps}
\end{center}
\end{figure}

For the case of an infinite $U$, the Eq.~(\ref{EQ:dGreen}) takes a simpler
form 
\begin{align}
G_{d\sigma} = \dfrac{1 - {\langle}{n}_{d\bar{\sigma}} \left( {\omega} \right){\rangle} + \displaystyle\sum_{q}\dfrac{V_{q\bar{\sigma}}{\langle}d^{\dagger}_{\bar{\sigma}}c_{q\bar{\sigma}}{\rangle}}{D_{2\sigma}\left({\omega},q\right)}}{ {\omega} - {\epsilon}_{d\sigma} - \alpha {\Sigma}_{0\sigma} - \displaystyle\sum_{k} \displaystyle\sum_{q}
\dfrac{V_{k\bar{\sigma}}^{*}V_{q\bar{\sigma}}{\langle}c^{\dagger}_{q\bar{\sigma}}c_{k\bar{\sigma}}{\rangle}}{D_{2\sigma}\left(\omega, k\right)}  } \;,
\label{EQ:dGreen1}
\end{align}      
where the zero-ordered self-energy is renormalized by a factor
\begin{equation}
\alpha= 1 +  \displaystyle\sum_{k} \dfrac{V_{k\bar{\sigma}}^{*}{\langle}c_{k\bar{\sigma}}d^{\dagger}_{\bar{\sigma}}{\rangle}}{D_{2\sigma}\left({\omega},k \right)} \;.
\end{equation}
Before we apply our formulation to discuss the Kondo resonance in the presence
of the spin accumulation, we examine various well-studied cases by numerically
solving Eq.~(\ref{EQ:dGreen1}). We choose the following parameters for our numerical 
calculation. The energy of the half-width of the impurity resonance in a 
nonmagnetic metal, $\Delta_{0}= - \mbox{Im} [\Sigma_{0\sigma}
(\omega+i0^{+})]$, is taken $200\;\mbox{meV}$ unless specified otherwise. 
The conduction band half-width is $D=100\Delta_{0}$. The degenerate impurity 
level is ${\epsilon}_{d} = -6 \Delta_{0}$. 

The simplest case is the conventional equilibrium
Kondo problem where neither impurity
state nor hybridization is spin-dependence. 
Figure~\ref{fig:fig1.eps} shows the impurity spectral density 
($\rho_{d\uparrow}=\rho_{d\downarrow}$) for four temperatures. 
As expected, the virtual bound level with a broad spectrum and a sharp
peak at the Fermi level known as the Kondo peak appears. The Kondo peak
is suppressed and broadened when the temperature is increased. 
The Kondo temperature which is defined as the full-width of the Kondo
peak is $T_{K}{\simeq}\exp[\pi {\epsilon}_{d}/2\Delta_{0}]$. These well-known 
results agree with many various approaches, e.g., the scaling 
analysis~\cite{Haldane} 
and the non-crossing approximation~\cite{Bickers}.
Meir {\em et al.}~\cite{Meir93} and 
Martinek {\em et al.}~\cite{Martinek1} has also used the EOM approach to
derive these Kondo peaks in the absence of spin polarization.

We now turn on the spin polarization. In a quantum dot, the spin polarization
is introduced via the coupling to a ferromagnetic lead. In this case, 
one can parametrize the spin dependence through the hybridization parameter,
i.e., $\Delta_{\sigma}=\Delta_{0}(1-\sigma P)$, where the parameter 
$0<P<1$ and $\sigma=\pm 1$ for spin up ($+1$) and down ($-1$). 
The results are shown in Fig.~\ref{fig:fig2.eps}. When $P$ is non-zero, 
the Kondo resonance splits [Fig.~\ref{fig:fig2.eps}(a)-(b)], i.e., 
the peak for spin up (down) spectral density shifts below (above)
the Fermi energy. Notice also that
the intensity of the peak for spin up (down) is suppressed (enhanced)
compared to the equilibrium Kondo peaks, see 
Fig.~\ref{fig:fig2.eps}(a)-(b) and Fig.~\ref{fig:fig1.eps}.    
These results are again consistent with those obtained by Martinek 
{\em et al.}~\cite{Martinek1}. Another way of introducing the spin-polarization
is to apply a magnetic field. In this case, the Zeeman splitting of the
impurity level becomes $\epsilon_{\sigma} = \epsilon_d + \sigma \mu_B B$, where $B$ is 
the magnetic field. We find that the Kondo peaks for spin up and down are
separated by $2\mu_B B$ (not shown), which agrees with 
that in Ref.~\onlinecite{Meir93}.

The agreement of our results with other approaches validates the EOM approach
of Eq.~(\ref{EQ:dGreen}) or Eq.~(\ref{EQ:dGreen1}). The significance of our EOM approach
is that it does not rely
on the additional renormalization introduced in the previous EOM technique
\cite{Meir91}. The purpose of the additional renormalization is to 
account for the spin dependent level splitting and broadening 
~\cite{Meir93,Martinek1}. The lack of rigorous justification for the existence
of the additional renormalization has cast a doubt for the effectiveness of
the EOM approach for the non-equilibrium Kondo problem. In our  
improved EOM formula, we have shown that the correct Kondo resonance can be
derived without introducing the additional renormalization.
Comparing with previous calculations, we have properly evaluated 
terms such as ${\langle}c^{\dagger}_{q\bar{\sigma}}c_{k\bar{\sigma}}
{\rangle}$, ${\langle}d^{\dagger}_{\bar{\sigma}}c_{k\bar{\sigma}}{\rangle}$ 
through Eq.~(\ref{EQ:Sigma1}). These terms make crucial contributions to 
the Kondo resonance splitting at low temperatures. Neglecting these terms
will lead to severe errors, which has to be recovered by the artificially adding an
additional renormalization. 

\begin{figure}[h]
\begin{center}
\includegraphics[width=8cm]{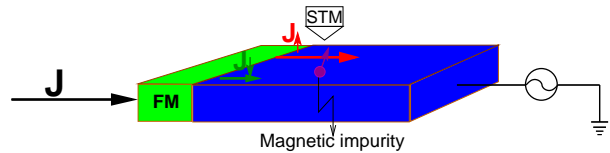}
\caption{(Color) Schematic illustration of a spin-current injected from a 
ferromagnetic layer to a non-magnetic conductor containing a magnetic impurity.
}
\label{fig:fig3.eps}
\end{center}
\end{figure}

We now apply the EOM equation to calculate the Kondo resonance in the presence
of the spin accumulation. We consider a bilayer
structure where a current flows from the ferromagnetic layer to the
non-magnetic layer containing Kondo impurities as schematically shown in
Fig.~\ref{fig:fig3.eps}. When a spin-polarized current injects into the
non-magnetic conductor, a spin accumulation is built-up near the interface.
Assuming that the ferromagnetic layer carries a spin-polarization of the 
current $P$, the spin accumulation is \cite{Johnson, Valet} 
$\delta m = (Pj \lambda \mu_B/eD) \exp(-x/\lambda)$ where $j$ is the 
current density, $\lambda$ is
the spin diffusion length, $\mu_B$ is Bohr magneton, $D$ is the diffusion
constant, and $x$ is the distance away from the interface.
If we only consider the Kondo impurity sufficiently close
to the interface, i.e., within the distance of $\lambda$, we can drop 
the spatial dependence
of the spin accumulation, i.e., $\delta m = Pj \lambda \mu_B/eD$.
\begin{figure}[t]
\begin{center}
\includegraphics[width=8cm]{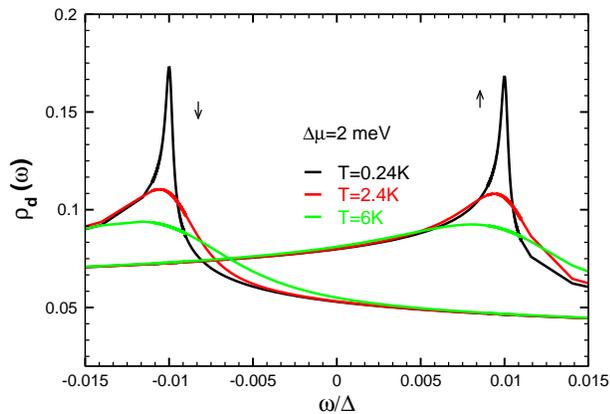}
\caption{(Color) Spectral density ${\rho}_{d\sigma}$ calculated via the EOM method for an infinity $U$ Anderson impurity for various temperatures in presence of spin accumulation with $\Delta\mu=2\;\mbox{meV}$.}
\label{fig:fig4.eps}
\end{center}
\end{figure}
To calculate the Kondo resonance from Eq.~(13), we specify the dependence of the
parameters on the spin accumulation. First, the spin accumulation makes the
chemical potential spin-dependent \cite{Johnson,Valet}. Specifically, 
the chemical potential splitting of spin up and down conduction
electron is $\mu^{\uparrow}-\mu^{\downarrow} = \delta m (eD\rho /\mu_B) 
= Pj\rho \lambda$ ~\cite{Qi}, where $\rho$ is the resistivity. 
Thus one should replace the Fermi level in Eq.~(13) by the spin-dependent
chemical potentials for the
spin up and down $\mu^{\uparrow}=E_F + Pj\rho \lambda/2 $ and
$\mu^{\downarrow}=E_F - Pj\rho \lambda/2 $, where $E_F$ is the Fermi level.
Second, the hybridization
parameter $\Delta$ would also be spin dependent since the density of states
of the conduction electrons is modified by the non-equilibrium electrons. 
However, the non-equilibrium electron density at very high current density
(say $10^7\;\mbox{A/cm}^2$) is at least several orders of magnitude smaller 
than the equilibrium electron density, and thus the correction to $\Delta$ is 
very small and we will assume that $\Delta$ remains spin-independent.
Finally, the spin accumulation could lead to the spin-dependent energy shift
of the impurity state. If one models the interaction between the
spin accumulation and the impurity via a phenomenological exchange
coupling, i.e., $H'=-J_{ex} \delta {\bf m}\cdot {\bf S}_i$, where $\delta 
{\bf m}$ is the spin accumulation and ${\bf S}_i$ is the impurity spin, the 
impurity level would be spin-split; this will be equivalent to the case
when the impurity is subject to
an effective magnetic field ${\bf B}_{eff}=J_{ex} \delta {\bf m}$. Although
the magnitude of the effective field could be respectable for a high current
density, the local level splitting by a magnetic field has already been
thoroughly investigated and thus we neglect the effect of the 
direct coupling between the spin accumulation and the impurity.
Therefore, we focus on the Kondo resonances due to spin dependent Fermi
levels. In Fig.~\ref{fig:fig4.eps}, we show the spectral density 
${\rho}_{d\sigma}$ for various temperatures with a spin-current-induced chemical potential shift. It is found that the spin-current influence 
on the Kondo resonance splitting is robust against temperatures. The more the amplitude of chemical potential relative shift by spin current, the more pronounced the Kondo resonance splitting between two spin channels will be. Noticeably, the amplitude of the split Kondo peaks  remains robust against the 
spin-induced chemical potential splitting (compare Fig.~\ref{fig:fig4.eps} with Fig.~\ref{fig:fig1.eps}), 
which is different from the case of a ferromagnetic metal as discussed before 
(see Fig.~\ref{fig:fig2.eps}(a)-(b)). This observation is the hallmark of the Kondo resonance splitting 
from the spin accumulation --- {a purely non-equilibrium effect.}

Advanced experimental techniques such as magnetic tunneling into quantum dots
and STM measurement would be able to detect the spin-split Kondo peaks. This
will provide a high resolution detection of the spin accumulation. 
For example, for a relatively small current density of the order of
$10^5$ A/cm$^2$, the Kondo peaks will be separated by about $\delta E =
0.02$ meV for a typical transition metal such as Fe (note that the width
of the Kondo resonance, which is of the order of the Kondo temperature).
In comparison, the reliable measurement of the spin accumulation based on the 
Silsbee-Johnson technique \cite{Johnson} 
usually requires the current density more
than $10^7$ A/cm$^2$ in order to derive sufficient large signals. 

In conclusion, we have developed an improved decoupling scheme for the
EOM approach in Kondo problems at low temperature. We show that by keeping
the previously discarded terms in the EOM approach, the additional 
renormalization is naturally implemented. We then apply the improved
EOM approach to study the Kondo effect in the presence of the spin accumulation.
We predict that the Kondo peaks are spin-split due to the spin dependence
of the chemical potentials. 

We are grateful to M. E.  Flatt{\'{e}} and Li Sheng for helpful 
discussions and suggestions. 
This work was supported by the Robert Welch Foundation No. E-1146 
at the University of Houston (Y.Q. and C.S.T.), 
by U.S. DOE under Contract No. DE-AC52-06NA25396 and under Grant Nos. 
LDRD-DR X9GT \& X9HH  (J.X.Z.), 
and by NSF-DMR-074182 (S.Z.) at the University of Missouri.

\bibliography{}

\end{document}